\documentclass[twocolumn,prl,showpacs]{revtex4}
\usepackage{amssymb}
\usepackage{amsmath,bm}
\usepackage{graphicx}
\usepackage[normalem]{ulem}
\usepackage{multirow}
\usepackage[utf8]{inputenc}
\usepackage[usenames, dvipsnames]{xcolor}
\usepackage{tensor}

\DeclareMathOperator\erfc{erfc}

\renewcommand{\sout}{\bgroup \color{red} \ULdepth=-.5ex \ULset}

\newcommand{\pdt}{$ N_\text{t}N_\text{p}/ N_\text{d}^2$}

\begin{document}

\title{Effects of QCD critical  point on light nuclei production}
\author{Kai-Jia Sun\footnote{%
Corresponding author: kjsun@tamu.edu}}
\affiliation{Cyclotron Institute and Department of Physics and Astronomy, Texas A\&M University, College Station, Texas 77843, USA}
\author{Feng Li\footnote{%
Corresponding author: fengli@lzu.edu.cn}}
\affiliation{School of Physical Science and Technology, Lanzhou University, Lanzhou, Gansu, 073000, China }
\author{Che Ming Ko\footnote{%
ko@comp.tamu.edu}}
\affiliation{Cyclotron Institute and Department of Physics and Astronomy, Texas A\&M University, College Station, Texas 77843, USA}

\date{\today}

\begin{abstract}
Using the nucleon coalescence model, which can naturally take into account the correlations in the nucleon density distribution, we study the effects of QCD critical point on  light nuclei production in relativistic heavy-ion collisions.  We find that the yield ratio \pdt~of proton ($p$), deuteron ($d$) and triton ($t$) increases monotonically with the nucleon density correlation length, which is expected to increase significantly near the critical point in the QCD phase diagram. Our study thus demonstrates that  the yield ratio \pdt~ can  be used as a sensitive probe of the QCD critical phenomenon. We further discuss the relation between the QCD phase transitions in heavy-ion collisions and the possible non-monotonic behavior of \pdt~in its collision energy dependence.
\end{abstract}

%\pacs{12.38.Mh, 5.75.Ld, 25.75.-q, 24.10.Lx}
\maketitle

\emph{Introduction.}
% 1.1 QCD phase diagram
According to the quantum chromodynamics (QCD), for a hadronic matter at sufficiently high temperatures and/or densities, the quarks and gluons inside the hadrons can be liberated to form a new phase of matter called the quark-gluon plasma (QGP)~\cite{Shuryak:1980tp}.   From calculations based on  the  lattice quantum chromodynamics (LQCD), the transition between the hadronic matter and the QGP is a smooth crossover if the matter has a low baryon chemical potential ($\mu_B$). This smooth phase transition has  been confirmed in experiments on ultrarelativistic heavy-ion collisions~\cite{Gyulassy:2004zy,Andronic:2017pug} in which the initially produced matter has a small $\mu_B$ and high temperature ($T$). Due to the fermion sign problem in LQCD~\cite{Gavai:2014ela}, it remains, however, an open question whether this smooth crossover changes to a first-order phase transition at high $\mu_B$ and low $T$~\cite{Stephanov:2004wx}, with a possible critical  endpoint (CEP), corresponding to a second-order phase transition, on the first-order phase transition line in the $\mu_B-T$ plane of the QCD phase diagram.   Besides its intrinsic interest, knowledge on the properties of QCD phase diagram at finite $\mu_B$ is also useful for understanding the structure of the inner core of a neutron star~\cite{Annala:2019puf} and the gravitational wave from  neutron-star mergers~\cite{TheLIGOScientific:2017qsa}.  

%1.2 Experiments
Locating the possible CEP and the phase boundary in the QCD phase diagram is one of the main goals for the heavy-ion collision experiments being carried out at the Relativistic Heavy Ion Collider (RHIC), the Facility for Antiproton and Ion Research (FAIR), the Nuclotron-based Ion Collider Facility (NICA), the High-Intensity Heavy Ion Accelerator Facility (HIAF), and the Japan Proton Accelerator Research Complex (J-PARC). For a recent review, see e.g. Ref.~\cite{Bzdak:2019pkr}.  By changing the beam energy ($\sqrt{s}$) in heavy-ion collisions, different regions in the $\mu_B-T$ plane of the QCD phase diagram can be explored.   Although at very high $\sqrt{s}$, the evolution trajectory of  produced matter in the QCD phase diagram only passes across the crossover line, it could move close to the CEP or  pass across the  first-order phase transition line as the $\sqrt{s}$ becomes lower.

%1.3 Observables
Since the hadronic matter, which is formed from the phase transition of the short-lived    QGP  created  in heavy-ion collisions, undergoes a relatively long expansion,    it is a great challenge to find observables that are sensitive to the phase transitions occurred during the earlier stage of the collisions.   Based on the generic feature of divergent correlation length ($\xi$) at the critical point of phase transitions and the resulting properties of self similarities and universality classes~\cite{Wilson:1973jj}, observables sensitive to the correlation length, such as the correlations and fluctuations of conserved charges 
~\cite{Hatta:2003wn,Stephanov:2008qz,Asakawa:2009aj,Stephanov:2011pb}, have been proposed. In particular, the fourth-order or kurtosis of net-proton multiplicity fluctuations has been suggested as   an  observable for the critical point because of its dependence on higher orders in $\xi$ and its non-monotonic behavior as a function of $\sqrt{s}$~\cite{Stephanov:2008qz,Stephanov:2011pb}.   However, the  event-by-event fluctuation in the number of net-protons  in these theoretical studies~\cite{Stephanov:2008qz,Stephanov:2011pb} is for protons and antiprotons in certain spatial volume,  which is very different from those within certain momentum cut that are measured in experiments~\cite{Adamczewski-Musch:2020slf,Adam:2020unf}. It is unclear how the measured fluctuation in   momentum space is related to the fluctuation in coordinate space due to the CEP, especially for heavy-ion collisions at low beam energies because of the lack of boost invariance~\cite{Asakawa:2019kek} and the effects of thermal smearing~\cite{Ohnishi:2016bdf}.
 
% 1.4 Light nuclei
On the other hand, light nuclei, such as the deuteron ($d$), triton ($^3$H or $t$), helium-3 ($^3$He), etc. that are produced in relativistic heavy-ion collisions, provide a promising tool to probe the spatial density fluctuation and correlation in the produced matter as they are formed from nucleons that are located in a very restricted volume  of $\Delta x\sim 2~ \text{fm}$ and $\Delta p\sim 100~\text{MeV}$ in phase 
space~\cite{Steinheimer:2012gc,Steinheimer:2013xxa,Sun:2017xrx,Sun:2018jhg}. It has been shown that    the fluctuation of nucleon density distributions can lead to an enhancement in the yield ratio $  N_\text{t}N_\text{p}/ N_\text{d}^2$ through the relation $N_\text{t}N_\text{p}/ N_\text{d}^2\approx \frac{1}{2\sqrt{3}}(1+\Delta \rho_n)$, where  $\Delta\rho_n$ is the average neutron density fluctuation in the coordinate space~\cite{Sun:2017xrx,Sun:2018jhg}.  Also, it has been proposed that this ratio could be enhanced because of the  modification of the  nucleon-nucleon potential near the CEP~\cite{Shuryak:2018lgd,Shuryak:2019ikv}.  Possible non-monotonic behavior of this ratio has indeed been seen in recent experimental data from heavy-ion collisions at both SPS energies~\cite{Sun:2017xrx} and RHIC BES energies~\cite{Zhang:2020ewj}.  However, no microscopic models have so far been able to  explain the data. 

% 1.5 Noval results
In this Letter, we point out that the yield ratio \pdt~in relativistic heavy-ion collisions is not only enhanced by a first-order phase transition in the produced matter, it also receives an additional enhancement from the long-range spatial correlation if the produced matter is near the CEP.  To quantify the dependence of this ratio  on the correlation length $\xi$ in the produced matter in heavy-ion collisions if its evolution trajectory passes close to the CEP of the QCD phase diagram, we derive an expression to relate these two quantities.   Because of the intrinsic resolution scale of around 2 fm given by the sizes of deuterons and tritons is comparable to the expected correlation length generated near the CEP~\cite{Berdnikov:1999ph}, the production of these nuclei is thus sensitive to the QCD critical fluctuations.  Our finding also suggests that the information on the CEP can be obtained from experiments on heavy-ion collisions by studying the collision energy dependence of the yield ratio  \pdt.

\emph{ Effects of critical point on light nuclei production in heavy-ion collisions.}
% 2.1 nucleon coalescence model
Although rarely produced in high energy heavy-ion collisions, light nuclei, such as $d$, $t$, $^3\text{He}$, helium-4 ($^4\text{He}$), hypertriton ($^3_\Lambda \text{H}$) and their antiparticles, have been observed in experiments at RHIC~\cite{Chen:2018tnh} and the LHC~\cite{Braun-Munzinger:2018hat}.   As shown in Refs.~\cite{Sun:2017xrx,Sun:2018jhg,Sun:2020pjz}, the effects of  nucleon density correlations and fluctuations on light nuclei production can be naturally studied by using the coalescence model~\cite{Sato:1981ez,Mrowczynski:1987oid,Scheibl:1998tk,Chen:2003ava,Sun:2017ooe,Blum:2019suo}.  In this model, the number of deuterons produced from a hadronic matter  can be calculated from the overlap of the proton and neutron joint distribution function $f_\text{np}({\bf x}_1,{\bf p}_1;{\bf x}_2,{\bf p}_2)$ in phase space with the deuteron's Wigner function $W_{\rm d}({\bf x},{\bf p})$, where  ${\bf x}\equiv({\bf x}_1-{\bf x}_2)/\sqrt{2}$ and ${\bf p}\equiv({\bf p}_1-{\bf p}_2)/{\sqrt{2}}$ are, respectively, the relative coordinates and momentum between its proton and neutron~\cite{Sun:2020pjz}, i.e., 
\begin{eqnarray}
N_\text{d}&=&g_\text{d}\int \text{d}^3{\bf x}_1  \text{d}^3{\bf p}_1  \text{d}^3{\bf x}_2 \text{d}^3{\bf p}_2 f_\text{np}({\bf x}_1,{\bf p}_1;{\bf x}_2,{\bf p}_2) \notag \\
&&\times W_\text{d}({\bf x},{\bf p}). \label{Eq:d01}
\end{eqnarray}
In the above, $g_d=3/4$ is the statistical factor for spin 1/2 proton and neutron to form a spin 1 deuteron.  As usually used in the coalescence model, we approximate the deuteron Wigner function $W_\text{d}$ by Gaussian functions in both $\mathbf x$ and $\mathbf p$, i.e., $W_\text{d}({\bf x},{\bf p})=8\exp\left({-\frac{x^2}{\sigma_d^2}}-\sigma_d^2 p^2\right)$, with the normalization condition   of $\int \text{d}^3{\bf x}\int \text{d}^3{\bf p}~W_\text{d}({\bf x},{\bf p})=(2\pi)^3$~\cite{Sun:2017xrx,Sun:2018jhg,Sun:2020pjz}. For the width parameter $\sigma_d$ in the deuteron Wigner function, it is related to the root-mean-square radius $r_d$ of deuteron by $\sigma_d  = \sqrt{4/3}~r_d\approx 2.26$ fm~ \cite{Ropke:2008qk,Sun:2017ooe}, which is much smaller than the size of the hot dense hadronic matter created in relativistic heavy-ion collisions.

% 2.2  coal: deuteron
For the proton and neutron joint distribution function $f_\text{np}({\bf x}_1,{\bf p}_1;{\bf x}_2,{\bf p}_2)$ in phase space, we take it to have the form   
\begin{eqnarray}
 f_\text{np}({\bf x}_1,{\bf p}_1;{\bf x}_2,{\bf p}_2)=\rho_\text{np}({\bf x}_1,{\bf x}_2)(2\pi mT)^{-3}~e^{-\frac{{\bf p}_1^2+{\bf p}_2^2}{2mT}},
 \label{Eq:fnp}
\end{eqnarray}
by assuming that protons and neutrons are emitted from a thermalized  source of temperature $T$ and  neutron and proton densities $\rho_{\rm n}({\bf x})$ and $\rho_{\rm p}({\bf x})$, respectively. In the above equation, $m$ is the nucleon mass and $\rho_\text{np}({\bf x}_1,{\bf x}_2)$ is the joint density distribution  function of protons and neutrons in the coordinate space, which  can be written as 
\begin{eqnarray}
\rho_\text{np}({\bf x}_1,{\bf x}_2) = \rho_{\rm n}({\bf x}_1)\rho_{\rm p}({\bf x}_2)+C_2({\bf x}_1,{\bf x}_2), 
\label{Eq:np}
\end{eqnarray}
in terms of the neutron and proton density correlation function $C_2({\bf x}_1,{\bf x}_2)$.    Substituting $f_\text{np}$ from Eq.~(\ref{Eq:fnp}) and Eq.~(\ref{Eq:np}) in Eq.~(\ref{Eq:d01}) and integrating over the nucleon momenta, we obtain the deuteron number as
\begin{eqnarray}
N_\text{d}
&\approx&N_\text{d}^{(0)}(1+C_\text{np})+ \frac{3}{2^{1/2}}\left(\frac{2\pi}{mT}\right)^{3/2}  \nonumber\\
&&\times \int \text{d}^3{\bf x}_1   \text{d}^3{\bf x}_2C_2({\bf x}_1,{\bf x}_2) \frac{e^{-\frac{({\bf x}_1-{\bf x}_2)^2}{2\sigma_d^2}}}{(2\pi \sigma_d^2)^{\frac{3}{2}}}.
\label{Eq:d03}
\end{eqnarray}
In the above, $N_\text{d}^{(0)}=\frac{3}{2^{1/2}}\left(\frac{2\pi}{mT}\right)^{3/2} N_{\rm p}\langle\rho_{\rm n}\rangle$, with $N_{\rm p}$ being the proton number in the emission source, denotes the deuteron number in the usual coalescence model studies without density fluctuations and correlations in the emission source, and $C_\text{np}=\langle\delta\rho_{\rm n}({\bf x})\delta\rho_{\rm p}({\bf x})\rangle /(\langle\rho_{\rm n}\rangle\langle\rho_{\rm p}\rangle)$ is the correlation between the neutron and proton density fluctuations, with $\langle\cdots\rangle$ denoting the average over the coordinate space ~\cite{Sun:2017xrx,Sun:2018jhg}.  In obtaining Eq.~(\ref{Eq:d03}), we have used the fact that the width parameter $\sigma_d$ in the deuteron Wigner function is much larger than the thermal wavelength of the nucleons in the emission source. We note that the  $C_{\rm np}$ term in Eq.~(\ref{Eq:d03}) for deuteron production has been carefully studied in Refs.~\cite{Sun:2017xrx,Sun:2018jhg,Sun:2020pjz}.

% 2.3 density correlation function
The nucleon density correlation  becomes important near the CEP, where it is dominated by its singular part given by~\cite{Wilson:1973jj,Koch:2008ia}
\begin{eqnarray}
C_2({\bf x}_1,{\bf x}_2) \approx \lambda \langle\rho_n\rangle\langle\rho_p\rangle\frac{e^{-|{\bf x}_1-{\bf x}_2|/\xi}}{|{\bf x}_1-{\bf x}_2|^{1+\eta}}.
\label{Eq:c2}
\end{eqnarray}
In the above, $\xi$ is the correlation length, $\eta$ is the critical exponent of anomalous dimension, and $\lambda$ is a parameter that varies smoothly with the temperature and the baryon chemical potential of the emission source.   The nucleon correlation length is similar to the correlation length  of the (net-)baryon density because nucleons carry most of the baryon charges in heavy-ion collisions.  In the non-linear sigma model~\cite{Stephanov:1998dy,Hatta:2003wn}, the correlation length is given by $\xi=1/m_\sigma$, where $m_\sigma$ denotes the in-medium mass of the sigma meson and decreases as the system approaches the CEP.  Since the value of the anomalous exponent $\eta\approx 0.04$ is small~\cite{Guida:1998bx}, it is neglected in the present study.  Using the fact that the nucleon number fluctuation  $\langle\delta N\rangle^2\propto \int \text{d}{\bf x} C_2({\bf x})\propto \lambda\xi^2$ near the critical point is always positive and enhanced, the $\lambda$ parameter is positive as well.

\begin{figure}[t]
  \centering
  \includegraphics[width=0.48\textwidth]{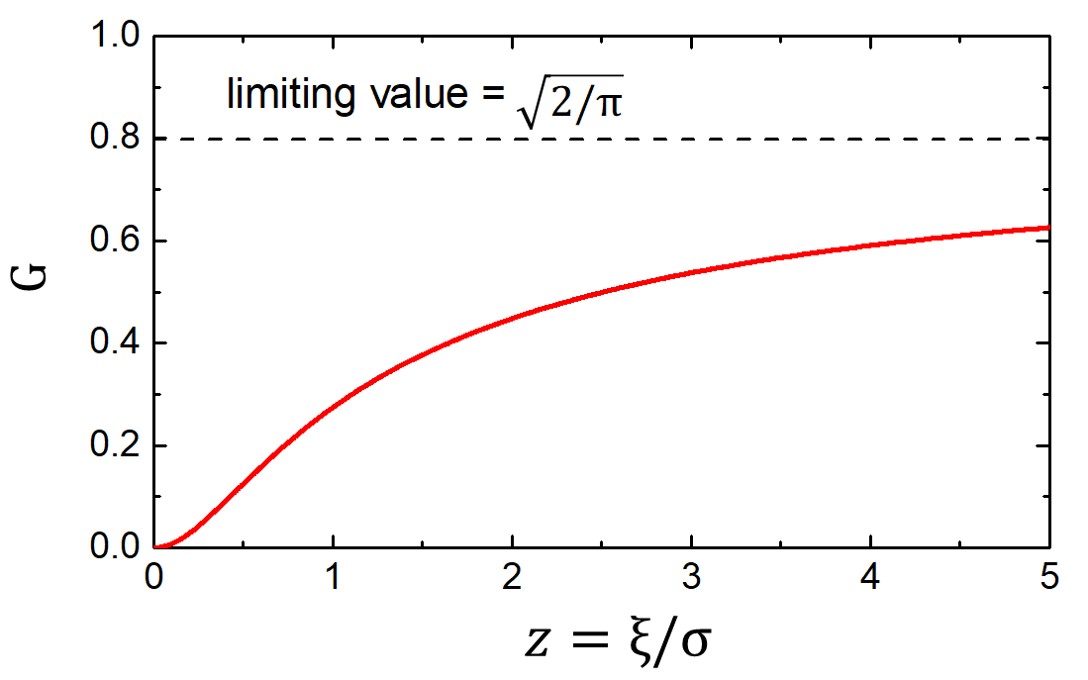}
  \caption{\protect  The   dependence of the function $G(\xi/\sigma)$  on the correlation length $\xi$ with $\sigma$ being the width parameter in the deuteron or triton Wigner function.}
  \label{pic:G}
\end{figure}

% 2.4 deuteron near CEP
With Eq.~(\ref{Eq:c2}), the deuteron number in Eq.~(\ref{Eq:d03}) then becomes
\begin{eqnarray}
N_\text{d}
&\approx&N_\text{d}^{(0)}\left[1+C_\text{np}+\frac{\lambda}{\sigma_d}G\left(\frac{\xi}{\sigma_d}\right)\right], 
\label{Eq:d04}
\end{eqnarray}
where the  function $G$ denotes the contribution from the long-range correlation between neutrons and protons, and it is given by
\begin{eqnarray}
G(z) %&=&\sqrt{\frac{2}{\pi}}\int \text{d}x xe^{-\frac{x}{s}-\frac{x^2}{2}} \notag \\
&=& \sqrt{\frac{2}{\pi}}-\frac{1}{z}e^{\frac{1}{2z^2}}\erfc\left(\frac{1}{\sqrt{2}z}\right), \label{Eq:G}
\end{eqnarray}
with $\erfc(z)$ being the complementary error function. The behavior of $G(z)$ is depicted in Fig.~\ref{pic:G}.  As the correlation length $\xi$ increases, the function $G(z)$ is seen to increase monotonically   and saturate  to the value $\sqrt{2/\pi}$ for $z\gg 1$ or $\xi\gg \sigma_d$. This means that the divergence of $\xi$   does not lead to a divergence of the deuteron yield, which is different from observables like the kurtosis of net-proton number distribution function. For small $\xi$, the function $G$ increases as $\xi^2$. At $\xi\sim \sigma_d$,  it increases linearly with $\xi$, and the increase becomes much slower for $\xi >3\sigma_d$.

% 2.5 triton near CEP
Similarly, the number of tritons from the coalescence of two neutrons and one proton is given by
\begin{eqnarray}
N_\text{t}&\approx&\frac{3^{3/2}}{4}\left(\frac{2\pi}{mT}\right)^{3}\int \text{d}^3{\bf{x_1}}   \text{d}^3{\bf{x_2}} \text{d}^3 {\bf{x_3}}\rho_\text{nnp}({\bf{x_1,x_2,x_3}})\nonumber\\
&&{\times} \frac{1}{3^{3/2}(\pi \sigma_t^2)^{3}}e^{-\frac{({\bf{x_1}}-{\bf{x_2}})^2}{2\sigma_t^2}-\frac{({\bf{x_1}}+{\bf{x_2}}-2{\bf{x_3}})^2}{6\sigma_t^2}},  \label{Eq:t01}
\end{eqnarray}
if the triton Wigner function is also approximated by Gaussian functions in the relative coordinates~\cite{Scheibl:1998tk,Chen:2003ava,Sun:2020pjz}. In the above, $\sigma_t$ is related to the root-mean-square radius $r_t$ of triton by $\sigma_t=r_t=1.59$ fm~{\cite{Ropke:2008qk,Scheibl:1998tk,Chen:2003ava,Sun:2020pjz}}. The three-nucleon joint density distribution function $\rho_\text{nnp}$ can be expressed as
\begin{eqnarray}
&&\rho_\text{nnp}({\bf x_1},{\bf x_2},{\bf x_3}) \approx \rho_{\rm n}({\bf x_1})\rho_{\rm n}({\bf x_2})\rho_{\rm p}({\bf x_3}) \notag \\
&&+C_2({\bf x_1},{\bf x_2})\rho_{\rm p}({\bf x_3})+C_2({\bf x_2},{\bf x_3})\rho_{\rm n}({\bf x_1})\notag \\
&&+C_2({\bf x_3},{\bf x_1})\rho_{\rm n}({\bf x_2})+C_3({\bf x_1},{\bf x_2},{\bf x_3}) \label{Eq:nnp}
\end{eqnarray}
in terms of three two-nucleon correlation functions $C_2({\bf x}_i,{\bf x}_j)$ with $i\ne j$, if one neglects the isospin dependence of two-nucleon correlation functions~\cite{Hatta:2003wn}, and the three-nucleon correlation function $C_3({\bf x_1},{\bf x_2},{\bf x_3})$. The contribution from the two-nucleon correlation functions to triton production can be similarly evaluated  as in Eq.~(\ref{Eq:d04}) for deuteron production.  Keeping only the leading-order term in the function $G$,  the triton number is then given by
\begin{eqnarray}
N_\text{t} &\approx&N_{\rm t}^{(0)}\left[1+\Delta\rho_n+2C_\text{np}+\frac{3 \lambda}{ \sigma_t}G\left(\frac{\xi}{\sigma_t}\right)+\mathcal O(G^2)\right],\notag\\
\label{Eq:t02}
\end{eqnarray}
where $N_{\rm t}^{(0)}= \frac{3^{3/2}}{4}\left(\frac{2\pi}{mT}\right)^{3} N_{\rm p}\langle\rho_{\rm n}\rangle^2$ is the triton number in the absence of nucleon density fluctuations and correlations in the emission source, and $\Delta\rho_n=\langle\delta\rho_n(x)^2\rangle/\langle\rho_n\rangle^2\ge 0$ is the relative neutron density fluctuation~\cite{Sun:2017xrx,Sun:2018jhg,Sun:2020pjz}. The term $\mathcal{O}(G^2)$ in    Eq.~(\ref{Eq:t02}) comes from the contribution of the three-nucleon correlation. We note that the $\Delta \rho_n$ is closely related to the second-order scaled density moment $y_2$ by $y_2  \equiv[\int \text{d}{\bf x}\rho_n({\bf{x}})][\int \text{d}{\bf x}\rho_n^3({\bf{x}})]/{[\int \text{d}{\bf x}\rho_n^2({\bf x})]^{2}}\approx 1+\Delta\rho_n $~\cite{Sun:2020pjz}, which has been frequently used to describe  the density fluctuation or inhomogeneity in coordinate space~\cite{Steinheimer:2012gc,Li:2016uvu,Sun:2020pjz}.  

% density in the y_2 is defined as a hydrodynamical variable.

% 2.6 double ratio: p-d-t
By considering the yield ratio \pdt, which can be considered as the double ratio of $N_{\rm t}/N_{\rm d}$ and $N_{\rm d}/N_{\rm p}$, one can eliminate the pre-factors of temperature and  the proton number  in Eqs. (\ref{Eq:d04}) and (\ref{Eq:t02}), which  depend on the beam energy and the collision system.  Neglecting the  difference between $\sigma_d$ and $\sigma_t$ for simplicity by denoting $\sigma\approx\sigma_\text{d}\approx \sigma_\text{t}$  and keeping only the leading-order term in $G$, the yield ratio \pdt~can be simplified   to
\begin{eqnarray}
 \frac{N_\text{t}N_\text{p}}{N_\text{d}^2} &\approx& \frac{1}{2\sqrt{3}}\left[1+\Delta \rho_n + \frac{\lambda}{\sigma}G\left(\frac{\xi}{\sigma}\right)\right].
\label{Eq:pdt}
\end{eqnarray}

%2.7  discussion
Eq.~(\ref{Eq:pdt}) shows that besides its enhancement by the neutron density fluctuation $\Delta\rho_n$~\cite{Sun:2017xrx,Sun:2018jhg}, the yield ratio \pdt~ is also enhanced by the  nucleon density correlations characterized by the correlation length $\xi$.  Although the density fluctuations $\Delta\rho_n$ in a  homogeneous system vanishes, the correlation length $\xi$ in the system becomes divergent if the system is close to the CEP.  On the other hand, in the presence of a first-order phase transition with the coexistence of two phases, the system could have large density inhomogeneity~\cite{Steinheimer:2012gc,Herold:2013qda,Li:2016uvu,Sun:2020pjz} and thus  non-vanishing density fluctuations, even though the correlation length in each phase is significantly smaller than in the case that the system is near the critical point. As a result, the existence of a first-order phase transition and the CEP in the system can both lead to enhancements of the yield ratio \pdt~if their effects can survive the hadronic evolution in a heavy-ion collision. 

% 2.8 heavier nucleus
The above result can be generalized to the yield ratios involving the heavier $^4$He ($\alpha$)~\cite{Sun:2017xrx,Shuryak:2018lgd,Shuryak:2019ikv}, e.g. the ratios
\begin{eqnarray}
 \frac{N_\alpha N_\text{p}}{N_{^3\text{He}}N_\text{d}} \approx  \frac{2\sqrt{2}}{9\sqrt{3}}\left[1+C_\text{np} + \Delta \rho_p  + \frac{2\lambda}{\sigma}G\left(\frac{\xi}{\sigma}\right)\right], \label{Eq:ratio1}  \\
 \frac{N_\alpha N_\text{t}N_\text{p}}{N_{^3\text{He}}N_\text{d}^3} \approx  \frac{1}{27\sqrt{2}}\left[1+C_\text{np} + 2\Delta \rho_n  + \frac{3\lambda}{\sigma}G\left(\frac{\xi}{\sigma}\right)\right].
\label{Eq:ratio2}
\end{eqnarray}
Compared to the yield ratio \pdt~in Eq.~(\ref{Eq:pdt}), these two yield ratios  show a larger sensitivity to the correlation length $\xi$. However, the yield of $\alpha$ particle in heavy-ion collisions is much more difficult to  measure precisely at high collision energies because of  its small value due to the large penalty factor $e^{-A(m-\mu_B)/T}$ for the yield of a nucleus with $A$ nucleons~\cite{Agakishiev:2011ib}.

% 2.9 advantages

As can be seen from Eq.~(\ref{Eq:pdt}), the yield ratio \pdt~encodes directly the spatial density fluctuations and correlations of nucleons due to, respectively, the first-order QGP to hadronic matter phase transition and the critical point in the produced matter from relativistic heavy-ion collisions. This is in contrast to the higher-order net-proton multiplicity fluctuations~\cite{Stephanov:2008qz,Stephanov:2011pb} based on the measurement of the event-by-event fluctuation of the net-proton number distribution in momentum space  with its relation to the spatial correlations due to the CEP still lacking~\cite{Asakawa:2019kek,Ohnishi:2016bdf}.  Also, the \pdt~ ratio has a natural resolution scale of around 2 fm, which is comparable to the correlation length $\xi\simeq $ 2-3 fm~\cite{Berdnikov:1999ph} that could be developed in  heavy-ion collisions.  Such an intrinsic scale is absent in the event-by-event fluctuation  of the net-proton number distribution  as it only depends on the selected rapidity range in the experiments~\cite{Ling:2015yau,Bzdak:2019pkr}.   

\emph{Collision energy dependence of the yield ratio \pdt.}
%  3.1 xi
Since both the density fluctuations and long-range correlations of nucleons in the emission source can lead to an enhanced yield ratio \pdt~ as shown in the above, the effects of QCD phase transitions can thus be studied in experiments from the collision energy dependence of this ratio. It has been  demonstrated in Ref.~\cite{Athanasiou:2010kw} that the correlation length $\xi$ along the chemical freeze-out line peaks near the CEP.  This indicates that the collision energy dependence of $\xi$ is likely to exhibit a peak structure with its maximum value at certain collision energy $\sqrt{s_H}$.  Due to the critical slowing down~\cite{Berdnikov:1999ph} in the growth of the correlation length, the value of $\xi$ is, however, limited to be around 2-3 fm  at the time the matter produced  in realistic heavy-ion collisions is near the CEP.  

% 3.2 y2
The neutron density fluctuation $\Delta\rho_n$  is mostly related to the first-order phase transition during which large density inhomogeneity could be developed due to the spinodal instability~\cite{Steinheimer:2012gc,Herold:2013qda,Li:2016uvu,Sun:2020pjz}. It was estimated in Ref.~\cite{Randrup:2009gp} that the largest effect of a first-order phase transition could be developed at around half the critical temperature~\cite{Randrup:2009gp} when the phase trajectory spends  the longest time  in the spinodal unstable region of the QCD phase diagram.  Therefore, the collision energy dependence of  $\Delta\rho_n$  is also expected to have a peak structure with its maximum value at a lower collision energy $\sqrt{s_L}$. 

% 3.3 p-d-t ratio
As a result, the yield ratio \pdt~  as a function of the collision energy $\sqrt{s}$ shows two possible non-monotonic behaviors.    The first one has a double-peak structure, with one peak at $\sqrt{s_H}$ due to the critical point and the other peak at a lower collision energy $\sqrt{s_L}$ due to the first-order phase transition. This double-peak  structure was conjectured in Ref.~\cite{Sun:2018jhg} without  the explicit relation between \pdt~ and $\xi$ given in Eq.~(\ref{Eq:pdt}).   Due to the flattening of the function $G$  given by Eq.~(\ref{Eq:G}) for large $\xi$, the signal from the critical point is broadened. It is thus also possible that  $\sqrt{s_L}$ and $\sqrt{s_H}$ are so close that the signals from the CEP and the first-order phase transition overlap, resulting in only one broad peak in the collision energy dependence of the yield ratio \pdt. 

%3.4  experimental results 
Possible non-monotonic behavior of the yield ratio \pdt~has been seen in recent experimental data from heavy-ion collisions  at both SPS energies~\cite{Anticic:2016ckv,Sun:2017xrx} and RHIC BES energies~\cite{Zhang:2020ewj}. In particular, the data for \pdt~at collision energies from $\sqrt{s_\text{NN}}=6.3~$GeV to $\sqrt{s_\text{NN}}=200~$GeV shows a possible double-peak structure. However, due to the large  error bars in the data, one cannot exclude the possibility that the double peaks are actually a single broad peak. To extract the collision energy dependence of $\Delta\rho_n$ and $\xi$ from the
experimental data by using Eq.~(\ref{Eq:pdt}), one needs to know the value of the $\lambda$ parameter in the critical region and its evolution during the hadronic evolution of heavy ion collisions, which requires studies that are beyond the scope of present paper.  

% 3.5 model calculation
In the statistical hadronization model, which assumes that both yields of  deuterons and trions remain constant during the hadronic evolution of heavy ion collisions, this ratio would increase with increasing  collision energy after  including the strong decay contribution to protons~\cite{Vovchenko:2020dmv}. Calculations based on transport models~\cite{Liu:2019nii,Sun:2020uoj} without the CEP in the QCD phase diagram all give an essentially energy-independent constant value for \pdt~ and thus fail to describe the data.   However, a recent multi-phase transport  model study~\cite{Sun:2020pjz}, which includes  a first-order QCD phase transition, shows that the density fluctuation or inhomogeneity induced during the first-order phase transition can  largely survive the hadronic evolution, because of the fast expansion of the produced matter, and eventually leads to an enhanced yield ratio \pdt~  at the kinetic freeze out of nucleons when they undergo their last  scatterings.  One thus expects the long-range correlation to similarly persist until kinetic freeze out and also lead to an enhancement of the yield ratio \pdt.

\emph{Summary  and outlook.} 
% 4.1 summary
Based on the nucleon coalescence model, we have obtained an explicit expression that relates  the yield ratio \pdt~ to the nucleon density correlation length $\xi$  in the hadronic matter produced in heavy-ion collisions, which could be appreciable if the produced matter is initially  close to the CEP in the QCD phase diagram. This ratio is found to increase monotonically with the dimensionless quantity $\xi/\sigma$ where $\sigma\approx 2$~fm denotes  the sizes of deuteron and triton.  This  enhancement is in addition to that due to  the large neutron density fluctuation $\Delta\rho_n$ that could be developed during a first-order QGP to hadronic matter  phase transition previously studied in Refs.~\cite{Sun:2017xrx,Sun:2018jhg,Sun:2020pjz}. Consequently, the collision energy dependence of this ratio is  expected to have a double-peak or a broad one-peak structure depending on the closeness in $\sqrt{s}$ between the signal of the CEP and that of the first-order phase transition.        Such a non-monotonic behavior in the collision energy dependence of the yield ratio \pdt~ has indeed been seen in the preliminary data from the STAR Collaboration.  Our study has thus led to the possibility of extracting  the information of the CEP and the phase boundary of QCD phase diagram from comparing the precisely measured data on the yields of light nuclei in heavy-ion collisions with those from theoretical models  based on  the transport approach~\cite{Sun:2020pjz} and the various  hydrodynamic  approaches~\cite{Murase:2016rhl,Singh:2018dpk,An:2019osr,Rajagopal:2019xwg,Nahrgang:2018afz}. 
 
\begin{acknowledgements}
This work was supported in part by the US Department of Energy under Contract No.DE-SC0015266 and the Welch Foundation under Grant No. A-1358.
\end{acknowledgements}

%\bibliography{myref}

\end{document}